\documentclass[]{svjour3}

\usepackage{amsmath}
\usepackage{hyperref}
\usepackage{longtable}
\usepackage{natbib}
\usepackage{graphicx}

\smartqed 
\journalname{   Experimental Astronomyo - DOI: 10.1007/s10686-011-9241-6}

\begin{document}

\title{Using Java for distributed computing in the Gaia satellite data processing}

\author{William O'Mullane$^{1}$ 
, Xavier Luri$^{2}$ 
, Paul Parsons$^{3}$ 
, Uwe Lammers$^{1}$ 
, John Hoar$^{1}$ 
, Jose Hernandez$^{1}$ 
}

\institute{$^1$European Space Astronomy Centre\\
$^2$Departament d'astronomia i meteorologia, Universitat of Barcelona, ICC-UB/IEEC \\
$^3$The Server Labs, Madrid, Spain}

\maketitle

\begin{abstract}
In recent years Java has matured to a stable easy-to-use language with the
flexibility of an interpreter (for reflection etc.) but the performance and
type checking  of a compiled language. When we started using Java for
astronomical applications around 1999 they were the first of their kind in
astronomy. Now a great deal of astronomy software is written in Java as are
many business applications. 

We discuss the current environment and trends
concerning the language and present an actual example of scientific
use of Java for high-performance distributed  computing: ESA's mission Gaia.
The Gaia scanning satellite will perform a galactic census of about 1000 
million objects in our galaxy. The Gaia community has chosen to write its 
processing software in Java. We explore the manifold reasons for choosing 
Java for this large science collaboration.

Gaia processing is numerically complex but highly distributable, some parts
being embarrassingly parallel. We describe the Gaia processing architecture
and its realisation in Java. We delve into the astrometric solution which is
the most advanced and most complex part of the processing. The Gaia simulator 
is also written in Java and is the most mature code in the system. This has 
been successfully running since about 2005 on the supercomputer "Marenostrum" 
in Barcelona. We relate experiences of using Java on a large shared machine.

Finally we discuss Java, including some of its problems, for scientific computing.
\end{abstract}

 \PACS{PACS 07.05.Kf   \and PACS 95.10.Jk  \and PACS 07.87.+v    }


\section{Introduction \label{sect:intro}}

\subsection{The Gaia Mission}
The Gaia satellite is destined for the Lagrange point L2 early in 2012 after launch 
summer 2012 from French Guiana aboard a Soyuz Fregat Rocket. Gaia is the European Space 
Agency's (ESA) sixth cornerstone mission. Its goal to make a phase space map of our galaxy. 
Spinning around its own axis in a Lissajous
orbit around L2 for five years gaia will continually scan the sky observing more than 
one thousand million ($10^{9})$ celestial sources, on average eighty times each. The
scientific goals of the experiment are manifold and covered in detail in \citep{gaia2001},
the data will help key research in the composition and formation of our galaxy.

Gaia contains two astrometric telescopes at a fixed angle of $106.5^{\circ}$ as well as a
radial velocity spectrograph and two photometers. The astrometric design 
allows true parallaxes (distances to stars) to be obtained \citep{1998A&A...340..309M}
 after careful data processing. A least-squares fitting scheme named the Astrometric
 Global Iterative Solution (AGIS) \citep{LL08} is currently foreseen to perform the astrometric
data reduction. Equally involved processing is required for photometry and spectroscopy.

Gaia processing software is being written by the Gaia Data Processing and Analysis Consortium
(DPAC). DPAC is a pan-European federation of institutes comprising of over four hundred 
astronomers, physicists and programmers. The consortium is led by eleven members of the 
community who form the DPAC Executive (DPACE).
DPAC not only has the responsibility to write the
processing software but also to run and maintain it until the final Gaia catalogue is produced
around the year 2020. The software will be run in six data processing centres - each 
responsible for a different facet of the overall processing. An overview of DPAC and the 
processing is provided in \citep{WOM+2007}.

\subsection{Computing}
Simulations of Gaia data have been in production at the University of Barcelona (UB) since about
1998. Some of these simulations require considerable computing power and are discussed in 
Section \ref{sect:simu}. Initial AGIS experiments
have been conducted using simulation data from UB in the past few years. 
The AGIS efforts are discussed in Section \ref{sect:agis}.
By nature the data processing must be distributed. If one considers that there are $\approx 10^{12}$ 
(low resolution cutout) images downlinked from Gaia, at one millisecond per image that is over 30 years of processor time.
Massive distribution is the only possibility to deal with this data. 
At this point the Gaia community has
quite some experience with large software and computing efforts, which is not unusual in the
science community. What some consider quite unusual is that all Gaia software is written
in the Java language. We discuss the choice of Java in Section \ref{sect:gaiajava}.

\section{AGIS - a complex computing problem}  \label{sect:agis}

The Astrometric Global Iterative Solution (AGIS) provides the rigid framework for all of
the Gaia measurements. Gaia is spinning freely and making observations which relate only 
to other observations made by Gaia. It is an absolute instrument - there is no
list of input stars such as there was for Hipparcos \citep{hip:catalogue}, the predecessor of Gaia. What this 
means is that the Gaia data must be reduced in a self-consistent manner such that all 
individual observations of celestial sources, the model of each source's position and motion, 
Gaia's own attitude, orbit and velocity must be in harmony. Later the entire system 
will be  aligned with the International Celestial Reference System (ICRS). 

The complete mathematical formulation for AGIS is presented in \citep{LL08} while the
computational framework is described in \citep{WOM2008}. Here we recap these aspects briefly
in section \ref{sect:agisrecap} before discussing results (section \ref{sect:agisresults}) and 
performance/manpower trade offs (section \ref{sect:agisman}).

\subsection{AGIS overview} \label{sect:agisrecap}
AGIS is a block-iterative solution for the Gaia astrometry. It consists of at least four blocks
which may be calculated independently. Each block is however dependent on the result of the
 other blocks. We term one step through all blocks an outer iteration, 
although the term outer is frequently dropped. 
This is simplistically formulated in the following equations :
\begin{align} 
S &= A + G + C  \label{eq:s} \\
A &= S + G + C  \label{eq:a}\\
G &= S + A + C  \label{eq:g}\\
C &= S + A + G  \label{eq:c}
\end{align}
\noindent which are discussed further below.

The vast majority of Gaia sources (Equation \ref{eq:s}) may
 be described by a six parameter astrometric model. These
parameters and their derivation are fairly standard in modern astrometry \citep{murray1983} 
and describe the  position and motion of the source in three dimensions. The parameters are :
\begin{table*}[!h]
\begin{tabular} {l p{0.7\textwidth}}
$\alpha$ & azimuthal angle of object a.k.a.\ right ascension\\
$\delta$ & angular distance from the celestial equator (north or south) a.k.a. declination\\
$\pi$ & annual parallax, the apparent shift in angular position when
viewed from opposite sides of the sun \\ $\mu_{\alpha*}$ & $\mu_{\alpha*} = \mu_{\alpha} cos (\delta) $ proper motion in $\alpha$ direction \\
$\mu_{\delta }$ & proper motion in $\delta$ direction\\
$\mu_{r }$ & radial velocity, motion in the line of sight direction  \\
\end{tabular}
\end{table*}
 
From the $\approx 80$~observations of each source made by Gaia a least-squares fitting of the 
observations to the model may be made. The fit must include several intricacies namely:
\begin{itemize}
\item the orientation of Gaia in space (or attitude, Equation \ref{eq:a})
\item the path of light through the instrument (or calibration, Equation \ref{eq:c})
\item global parameters such as relativistic numbers (Section \ref{eq:g})
\item light bending according to general relativity (see \citep{klioner2003})
\end{itemize}

Current best estimates for  attitude, calibration and global parameters are used for any given
fit of the source. This fit must be performed on all sources but to make the
problem tractable a subset of sources may be treated first. Around $10^{8}$ sources
are needed to make the global reference frame but it could be as many as
$5\cdot10^{8}$. 
Once this solution has converged the source 
calculation (using the final attitude etc.) may be performed on all remaining sources. 
This is an important efficiency improvement.
A direct solution for a large number of sources  has been
shown to be intractable  \citep{AB2008}.

The application of the source equation (known as source update) requires the gathering of
all observation of the source and the current attitude, calibration and global parameters. The 
application of the relativistic model requires precise ephemerides of the planets and of the 
satellite. Our current approach is to group observations together on disk for easy loading, 
since there is no cross talk between the source equations i.e.\ all sources may be processed 
in parallel if we wish. The other data required is sufficiently small (order of a few hundred 
MB) that it may be loaded
once in memory and used to process multiple sources, typically batches of several thousand
sources per job on a processor. 

Attitude and calibration on the other hand require all observations in given time periods. 
We do not need to hold on to the observations but may add their contributions to partial
equations. Hence as each source is updated the updated parameters and the observations 
are used to update the partial solutions for the attitude and calibration. The system has been run with up to 
100 computing nodes and 1400 Java threads. We constantly encounter bottlenecks usually in the
calibration and attitude processes for the obvious reason that they need to see all sources. The processing
of several thousand sources in a job makes the communication more efficient - we send fewer large messages
rather than millions of small ones.

\subsection{Results} \label{sect:agisresults}
AGIS has been running at ESAC since early 2006 with increasing numbers of sources and observations.
This is discussed briefly in \citep{WOM2008} from which we reproduce table \ref{tab:agistimes} 
below. Since the entire system is designed for AGIS and we have few interconnects between
processes we did not find it necessary to use any HPC library nor GRID software. Effectively
jobs are taken from a {\em Whiteboard} by numerous {\em DataTrains} which load the 
appropriate  data and process it writing the results back to a database. 
The {\em Whiteboard} may simply be considered as a table in a database where jobs are written an updated.
{\em DataTrain} is a term coined within DPAC for a process which runs through a set of data passing
it to a series of algorithms. It provides an abstraction layer between algorithm and data access.

The machine used  for the Gaia tests so far has been purchased in installments by
 ESA. The initial machine for the first tests in 2005 consisted of twelve nodes
 each with 6 GB of RAM and two processors (3.6GHz Xeon EM64T). An EMC storage area
 network (SAN) with 5 TB
 of disk is attached to the nodes using fibre optic cable and the nodes are also
 connected via Gigabit Ethernet cards in a local area network.  AGIS has also
 been tested, and shown to run as fast, using a cheap Rack Server Network Attached
 Storage device of 6 TB.
 This cluster was upgraded to 18 nodes in 2006 and a further 4 nodes but with
 quad core processors were added in 2007. The quad core processors functioned very well
 for us providing the performance of 4 processors and in table \ref{tab:agistimes} the number 
 of processors counts each core as a processor.  
 
 This is not very special machine, it consists of  standard Dell power edge
 blades with standard Intel Xeon processors running Red Hat Linux.  The blades
 are housed in a standard Dell rack. There is no special HPC software
 used, a network of normal Linux machines could work in the same manner. As such this
machine perhaps does not merit the term HPC but we will move toward 10 Teraflops in the
coming years.

\begin{table}
\caption{AGIS run times decrease as more processors are added. Note that the
data volume increased from 2005 to 2006 from 18 months to 5 years, the processor
power also increased but the run time went up. This was dramatically improved in
2007. In 2010 we have implemented a new math frame work around Conjugate gradient - this provides a better solution at some cost. 
The normalised column shows throughput per processor in the system (total
observations/processors/hours) e.g. an indication of the real performance.}
\label{tab:agistimes}
\begin {center}
\begin{tabular}{|l|l|r|r|r|}\hline
{\bf Date } &
{\bf Observations } &
{\bf Procs } &
{\bf Time  } &
{\bf Normalised }
\\\hline
2005 & $1.6 x 10^7$ src & 12 & 3h & $0.9 x10^6$ 
obs/hour\\\hline
2006 & $8 x 10^7$  & 36 & 5h & $0.5 x 10^6$obs/hour\\\hline
2007 & $8 x 10^7$  & 24 & 3h & $1.3 x 10^6$ obs/hour\\\hline
2008 & $8 x 10^7$  & 31 & 1h & $3.2 x 10^6$ obs/hour\\\hline
2009 & $2.6x10^8$  & 50 & 1.8h & $2.8 x 10^6$ obs/hour\\\hline
2010 & $4x10^9$  & 68 & 9.5 &  $6.2 x 10^6$ obs/hour\\\hline 
\end{tabular}
\end{center}
\end{table}

Various different types of astronomical test have been performed to show AGIS produces the
correct solution for Gaia. For this paper we are mainly interested in performance however, which
is best presented by table \ref{tab:agistimes}.
We see that AGIS has been slow in 2007  but finally it was made much faster in 2008 with a 
prospect of more improvement to come. It should be noted that more functionality has also been 
introduced each year - a more complete source model, a more complex calibration scheme etc.\ but
the efficiency has been increased. 
 Some speedups are due to profiling while others are due to 
mathematical methods. Between the Lund group and the ESAC group finally  the problem is being tackled from both
math and computing side with competence. Early work on AGIS was not so fruitful as it treated
the problem as something to just be {\em run} on a computer rather than a problem for which a 
system needed to be designed to make it efficient.  In that system math and computing 
techniques have been questioned and varied to arrive at the efficient implementation we have today.

\subsection{Performance vs Manpower} \label{sect:agisman}
We know we could write some parts of AGIS to run perhaps two or more times faster than Java. 
We could get a little more performance by utilising special machines and rewriting our code 
again for example toward a cell processor hybrid such as the Roadrunner. Speed of code is indeed 
an issue for Gaia in general but manpower is perhaps a bigger issue - we have a great deal of 
processing software to write. The estimated manpower in man years for the Gaia processing is around
2000. Granted this includes operation but development alone pre launch is estimated at 1000 
man years.  In the case of the astrometric solution we have good records. The initial working
AGIS in ESAC was done in about 180 man days so just over half a man year. It was clear to
us that this was only possible in Java - in C or C++ this would have taken far longer. This claim
is of course difficult to quantify, as a programming collective we have decades of experience with 
C and C++ projects and it is our opinion that coding in Java is cleaner and faster. 
Even in Java the problem is difficult. The previous solution which AGIS replaced was worked on 
for four years by various people; to be modest lets say only 8 man years went into it, but it
was more like 12. To date on AGIS we have spent about 15 man years and we estimate to 
finish AGIS in Java we need another 10 man years with probably more after launch. There
is also post launch maintenance to consider - we feel Java maintenance will be lower
cost than say C, and we have eight years of operations to consider - we have 27 man years in 
the plan for this. 

Of course it is difficult to calculate but we would need far more manpower to
manage code in C and more still if we were to customise for a particular computer 
system (especially as it would become obsolete during our operations). Again just for AGIS
we are talking of over 50 man years of effort which in today's monetary terms is around 
5 million euro. Even 20\% more effort to code in  another language 
would be a considerable sum. We think we can buy the hardware we need to run AGIS
for around a million euro. It has been observed on other space programs than manpower
typically ends up costing more than machines and apparently we are no different. 
Our intention is also to buy the most common and cheap processors - these are the the chips
Java is usually running fastest on as well because of their ubiquity. Hence a special more
super computer oriented machine would probably cost more. Finally there is power consumption
 if we could make AGIS 4 times faster we could save 25\% on power. Energy costs may well 
go up but probably not faster than manpower costs since the saving on energy requires
spending on manpower to customise code for our "hard to code" problem these will cancel each other out.


\section{Gaia Simulations - use of shared computing resource} \label{sect:simu}

The development of the \textit{Gaia simulator} has been an integral part of 
the Gaia data processing. This software tool is designed to provide realistic simulations
of the Gaia observations, closely mimicking the format and content of the data
that Gaia will send to ground. The main purpose of this simulated data has been
the feeding of the data processing chain in order to test and validate it, although
it has also been used for other purposes, for instance to evaluate some satellite
design options and to prepare the mission scientific case.

As mentioned above, the simulator has been running at the University of Barcelona
since 1998. Its initial development was closely tied to the first attempts
to develop a viable \textit{Global Iterative Solution} but now it has become a mature 
tool able to simulate a wide variety of celestial objects, physical and instrumental 
effects and data formats for the multiple data processing modules developed
by the Gaia DPAC.

In this section we will first review our experience with the development in Java,
the advantages and drawbacks we found, and then discuss a specific (and extreme) 
example of Java versus C performance found during this development.

\subsection{Java, a new kid on the block}

When the development of the simulator was started the first choice to be made
was the programming language to use. The team undertaking this task was at the 
time (1998) mainly (and almost exclusively) composed of scientists, with few
software engineering expertise. The obvious choice, based on the programming
experience of the team, would have been FORTRAN with perhaps C as a second,
but somewhat frowned upon, choice. Furthermore, astronomical and numerical
libraries were available in FORTRAN, and in most cases also in C, but not in
Java. 

However, Java was finally chosen for two main reasons:

\begin{enumerate}
 \item The ongoing \textit{Global Iterative Solution} development contract
       was going to be implemented in Java by requirement of ESA. Given that
       the two projects were closely tied at the time using a common language
       was natural.
 \item The advice of professional software engineers from ESA and some members
       of the team with software engineering background pointed towards
       the use of an Object Oriented language, strongly advising against FORTRAN
       for a large collaborative project. This left the choice between C++ and
       Java.
\end{enumerate}

At the end Java was chosen, but not without frequent second thoughts in the coming
years. The widening scientific community that was becoming involved in the project
was reticent for some time to adopt the new language, specially given the accumulated
experience in FORTRAN and C and the lack of libraries in Java.

However, eleven years later the landscape has much changed, and now Java has 
been even more widely (almost completely) adopted as a viable programming
language for scientific programming in the Gaia simulator community. Furthermore, 
the project has become large and complex, involving development teams distributed
around Europe. The management of the project has at this stage adopted many
professional software development tools (UML, Hudson, PMD, Checkstyle, etc.) that
would not have been available had a language like FORTRAN been chosen, making 
the coding much more robust and reliable. Also, like the team developing 
AGIS and based on previous experiences, they feel that the development has 
been quicker and more seamless than it would have been using C++. The motives
for this feeling are varied, but one of the big reasons is that thanks to the
garbage collection implemented in the JVMs memory leaks are less of an issue.
Since the simulator is quite intensive in memory usage, not having to worry about 
this problem is seen as a great advantage.

In short, the initial reticences on the use of Java for scientific programming in Gaia 
have vanished, to the point that one of the managers of the simulator
development has changed the teaching of programming for first year physics 
students from FORTRAN to Java.

\subsection{The blessing of portability}

One of the advantages of Java not mentioned in the previous section is its
portability. As said, the simulator has been running for eleven years and
during this time it has been ported to several machines, and in the process
Java has shown that its portability is real and practical. 

Due to its intensive computation needs the Gaia simulator has mostly been 
run at supercomputing centers, using significant amounts of computation time
It was initially run at the \textit{Centre de Supercomputaci\'o de Catalunya} 
(CESCA) were several medium-sized clusters were used: an IBM SP2, a Compaq
HPC320, an HP RX2600 and an SGI ALTIX 3700. In all cases the migration from
one cluster to the next was seamless, the only serious complication being
the adaptation of the execution to the different queue systems in each 
machine. Later on, in 2005, the simulator was moved to the \textit{Barcelona
Supercomputing Center} (BSC) where the \textit{Mare Nostrum} supercomputer had
been just installed. Again, the migration was quite simple, with the major
complication being how to adapt the execution to the new distributed
environment and queue system. Today the simulator is still running there,
having cruised through a \textit{Mare Nostrum} upgrade in 2006 without
needing any significant change and having consumed a some millions of CPU
hours and generated many terabytes of simulated data. 

In these years the portability of Java has therefore been a major advantage,
saving time and resources that otherwise would have been devoted to
adapting the code to the new environment. Furthermore, the portability
of Java has also allowed the running of the simulator on other environments
like local clusters for testing and development and a cluster at CNES
for tailored small simulations through a web page manager.

\subsection{Performance in scientific computation}

In the initial years of the project the performance in numerical computing 
of Java was discussed many times. Not having previous experience on its
use for scientific computation there was some worry that the language
could prove to be too slow at some point of the development. However,
some initial tests with a set of some often-used numerical algorithms
showed that Java was not much slower that C in solving these problems, 
thanks to the \textit{Just In Time} (JIT) compilation in the Java
virtual machines.


In the next years the JIT virtual machines steadily improved, and the
first working code was developed in the project, showing in practice 
that the implementation in Java was not producing any of the feared
bottlenecks. Nowadays, Java has become part of the landscape 
in scientific programming in Gaia, and although worries about performance 
are still occasionally an element of resistance, given the accumulated 
experience and clear savings in development time performance is no longer an 
issue.

Furthermore, the development model in DPAC is based on six-month development
cycles; at the end of each cycle an improved version is produced and
tested, leading to further requirements, corrections and improvements for
the next cycle. In this framework the optimisation of the system is not as
important as the maintainability and flexibility of the code to allow
such a quick production. The robustness and clarity of Java 
helps a lot in such a process, and as mentioned before has probably saved
a significant amount of money in programming resources. 

A real example encountered during the simulator will illustrate this point. 
A key piece of the Gaia astrometry is the calculation of the so-called ``relativistic corrections'', 
the changes on the apparent position of the objects in the sky deriving from relativistic effects: aberration, 
gravitational light bending, etc. This is a complex calculation taking into account the ephemeris of the major 
solar system bodies and requiring (for a $\mu$as accuracy) to reach the limit of the numerical 
precision of variables of the ``double'' type (64 bit floating point).

An initial (legacy) implementation of these calculations was available from S. Klioner 
in the form of C code and was used in the simulator until 2008 through JNI calls.  
At that time, in order to avoid the inconveniences of mixing two languages in the simulator, 
the same author developed for DPAC a new implementation coded in Java. 
Both implementations were thoroughly compared and results matched at sub-$\mu$as level. 
However, during the testing it was found that the computation times differed substantially 
between the two versions, the Java version being between 4 and 10 times {\bf faster} than 
the C version! Possible external sources of this difference (like overheads of the JNI calls)
 were ruled out and it was concluded that the difference was actually intrinsic to the code.

It is possible, even likely, that an optimisation of the C code would make it much more 
efficient, to at least the level of the Java code. However, this example clearly shows how 
the same developer did a quicker and better job in Java (a language that, unlike C, he was unfamiliar with). 
On the other hand, the difference possibly comes also from the excellent refactoring afforded by 
the new JITs virtual machines that now automatically makes many of the performance fixes which
 previously had to be manually implemented - in C of course all optimisations must be done by the coder.

Finally, the optimization work on the Java code has continued. The team is now 
exploring the increasing possibilities of the options for aggressive optimization and 
garbage collection tailoring available in some Java virtual machines (specially the IBM one, 
used in the MareNostrum supercomputer) that will possibly lead to further improvements 
in the performance of the simulator.

\section{Java, Gaia and commercial uses} \label{sect:gaiajava}
When \citep{2000ASPC..216..419O} was presented at the Astronomical Data Analysis and Software
Systems (ADASS) conference in 1999 most people in that community did not know what 
Java was and this was the only paper mentioning Java. 
Two years later about half of the astronomy related projects at ADASS involved Java.
Some project must take the first step, in this case Gaia and Planck were experimenting
with Java. These experiments possibly made it easier for the Integral SOC and the  
entire Herschel Science Ground Segment to be written in Java \citep{2004ASPC..314..376W}.
Gaia and Java go back many years, the initial prototype for the Global Iterative Solution
was already in Java in 1999 \citep{1999BaltA...8...57O}.

Java in the late 90s was not the same as it is today but already then it was seen 
to have potential for science development. For a project like Gaia we were faced with 
a life span of over 20 years for our software and an entrenched Fortran community. 
It was clear already back then, as discussed earlier, 
that Gaia processing software would not be written in 
Fortran but needed to be in a more modern, preferably OO, language, C++ being the obvious 
choice. The transition for some scientists from Fortran to C++ was seen as difficult 
or impossible. Java simply worked more easily and was chosen and eventually accepted by all.

\subsection{Negative Aspects}
We do not claim Java is the silver bullet \citep{10.1109/MC.1987.1663532} and 
we have several problems which we live with especially in numerical coding. As already
mentioned especially when we started, there were no good numerical libraries.
The situation improved from about 2000 onwards when Java started to become really usable for
scientific computing in terms of robustness and performance. A number of promising
numerical library developments took place (e.g.\ JAMA - A Java Matrix Package) but
until around 2005 most of them had stopped, leaving the distributions and code available
but largely unmaintained. A positive exception is the Apache Common Math library which is
still in active development and in extensive use in almost all of Gaia's data processing
software. Early versions of Apache Common Math were missing linear algebra functionality needed by
us but that has improved with time.

In parallel with diminishing efforts for developing numerical libraries we observed
a general decline in support for Numerics in Java. For instance the
Numerics Working Group of the Java Grande Forum (\footnote{\url{http://math.nist.gov/javanumerics/}}),
initially a driving force behind many positive developments around Java Numerics,
has effectively ceased to exist. This is a bit worrying along with a general perceived lack
of support for Java in the traditional conservative supercomputing scene that in terms of
languages remains to be dominated by Fortran and C until today.

In 1998 W.~Kahan, one of the key persons behind the IEEE 754 Standard for Binary Floating-Point Arithmetic,
delivered a keynote ``How Java's Floating-Point Hurts Everyone Everywhere''
\cite{Kahan1998} at a Java HPC workshop. In this contribution Java gets harshly criticized for
for a number of IEEE non-compliances which could and should have been avoided by the language
designers. In subsequent years Sun was repeatedly asked to rectify the known deficiencies but chose
to ignore all complaints and even today those issues raised more than 10 years ago are, to our
knowledge, still present in all existing Java implementations. With Java now under Oracle's control
it seems unlikely that the situation will change in the foreseeable future.
Fortunately, in our view most of the points are fairly subtle in nature and are unlikely
to show up as perceivable flaws in ``ordinary'' numerical application code.
We can confirm this for all of Gaia's software and in particular for the AGIS system
described here in more detail (Sect.~\ref{sect:agis}) with one exception: Java does not provide any
means to use what IEEE 754 calls 'trap-handlers' for capturing and dealing with numerical exceptions
like division by zero. Every arithmetic operation in Java always delivers a valid result
and that can make the debugging of numerical code extremely difficult and time consuming.
As an example, there was a coding error in the AGIS attitude update that caused the
introduction of an NaN value into a large matrix of equations which then propagated and spread through
the matrix before eventually leading to NaN in an end result much later. Without the option
to have an exception raised at the first occurrence of an NaN condition the only way to
find the problem is to check explicitly every intermediate result for NaN (Double.isNaN(x)).
For obvious reasons this is not a viable option in a large numerical code.

%

\subsection{Features}
One of the things which made java work well for us was its built in features. We
were especially happy to have multi threading and distributed programming such as RMI 
(Remote Method Invocation - allows calling a method on an object on another machine)
built in to the language. 

\subsection{Rise of Java in industry }
At the same time (back in the late 90's) we saw in industry a surge of Java
programming as it pushed from the the Web language clearly into the back office.
Enterprise Java Beans were starting to appear behind web pages accessing
databases and wrapping legacy applications to make them network available. Java
was no longer an "applet" language for making more interactive web pages - it was
handling credit cards and transactions and doing serious work. The fact that
Java is 100\% portable to all O/Ss where the JVM has been ported to and
backward compatible means that using libraries is no longer a painful issue.

Companies such as IBM, Oracle, BEA, Sun and many open source vendors have
created Java application servers that support the Java standards in JavaEE
produced by Sun.

Since then frameworks to further aid productivity in the development of
Java software such as Spring and Hibernate have appeared on the scene and these
have helped with the adoption of Java as the technology of choice for
developing software. In addition the fact that there are thousands of open
source libraries available to use in Java projects has also helped the spread
of Java.

\subsection{Maintainability, Robustness and Performance }

When developing software there are a number of key issues that have to be
addressed including but not limited to
Maintainability,
Robustness,
Scalability

Addressing maintainability first, Java compared to other languages such as C++
and Fortran offers a number of advantages. The defacto editor for Java, Eclipse,
provides a number of features that makes creating and maintaining Java code
easier; Graphical syntax highlighting, Refactoring wizards to improve the
design of the application; Plugins to produce documentation. Java was also one
of the first languages to have a really good implementation of the xUnit testing
frameworks with the widely adopted JUnit. With Java 1.5 and the introduction of
Generics and Annotations and with techniques such as Aspect Oriented Programming 
(AOP - not currently used on Gaia)
 we have the possibility of developing software with less code and the less code there is,
the more maintainable it is.

Java is more robust than C++, primarily because the C++ component of all Java
programs, the JVM itself, is the same for all Java programs and therefore tried
and tested millions of times.

Many people believe that Java is not as scalable as other languages because of
the overhead of the Java bytecode interpretation. However the Java Hotspot
server compiler, especially the one in Java 1.6, is incredibly efficient at
optimising code that is called very frequently.

\subsection{Portability and the Cloud}
Although there are minor problems with Java portability this is usually in Graphical User 
Interfaces rather than in the type of code forming the majority of the Gaia environment. 
As stated in Section \ref{sect:simu} our simulation code has been ported over many years 
to many platforms with little effort. Currently development is done on MacOs, Linux and windows 
systems without issue. Such portability lead us to consider using the Amazon
E2C for the Astrometric solution  (Section \ref{sect:agis}). We, unlike CERN \cite{Shiers2009559} 
are not yet tied in to one architecture. Although these days even CERN use some cloud resource. 

 With about 2 person weeks of effort part of our team (Parsons and Olias) got
AGIS running on the cloud. Some problems were encountered with the database configuration,
although at least Amazon already hand a VM with Oracle on it. Also a scalability problem
in our own code was found and remedied - prior to this we had no opportunity to 
run 1400 threads. 
It worked well, at least as well as our cheap in house cluster. Hence we agree with the
detractors of the Cloud - it is not a {\em supercomputer} with super fast interconnect, but
then one is not paying super prices either. 

In fact although our intention is/was to build a cheap cluster (around 1MEuro) for AGIS 
we estimate all AGIS mission processing for 100 Million Stars could be done for 
about 400KEuro. When energy is factored in this makes Amazon look very attractive.
It largely depends on idle time however - and we in any case would not go without an in 
house cluster for testing and development. It also appears we now may need to process
500 Million sources to counteract possible spatial error correlations (as presented by Berry Holl
in Heidelberg in 2009). This would bring the saving close to zero - but one must question if 
the likes of Amazon can do a cheaper job of maintaining a cluster.

The availability of a much larger number of nodes than
we can buy is very interesting both for testing and for production. By using Amazon 
we could perhaps do our processing faster by using more nodes and still have it cost less
than an in house machine. We shall continue to experiment and evaluate this option - in 
any case the final machine at ESAC for the processing will not be purchased until we are
a couple of years into the mission around 2015.
  
\subsection {Virtualisation - non uniqueness} 
The real power of Java comes from the notion of running on a virtual machine. 
This should not be, but often is, confused with interpretation such as done in LISP and
Smalltalk. This virtue is not shared by many new languages or schemes perhaps the most
all encompassing of which is the Microsoft .NET framework in which many languages are compiled
to the same virtual machine. We are not trying to say Java is the only language with this
feature. In fact with  virtualisation suddenly any particular machine/OS/langugae configuration 
can be equally portable with its own virtual machine. Suddenly putting those legacy apps in a cloud
may not be such a hard decision. 

\subsection{ Future of Java } 

The rise of Java in industry and positive experience we had we Java compared to
C++ reinforced our choice of Java as the language for Gaia. This choice was
reaffirmed within the Gaia community in 2006 \citep{LL:JH-002}. Indeed at the
time of writing all science development missions at ESAC are using Java as their
programming language.

We predict that Java will be the language of choice for the foreseeable future
because of all the advantages outlined earlier, although we believe that within
the Java Virtual Machine, we will see more use of dynamic languages such as
Ruby and Groovy for areas of the application that will need to be changed very
frequently.

\section{Conclusion} \label{sect:conc}

One may argue about the definition of High Performance Computing but within the Gaia
project since the late 90s we have certainly be doing numerically intensive computing 
and handling increasingly large amounts of data in a highly distributed manner. 
All of this is done using Java. 
We argue that this is a good option for long running projects where portability and 
maintenance are possibly more important than squeezing the last FLOP from a 
specific processor. 

The portability of Java has served us well in the last decade allowing code to move from
SPARC station to Linux boxes, WinTel and even the Marenostrum super computer. Of late
we have also leveraged Amazon's E2C resources.

We are not alone in the astronomy world, many other projects are using Java or other 
high level languages.  The ESA missions Herschel, Planck and 
many archives are Java users. Others such as Spitzer and  JWST mentioned using Java at
least in part. Other higher level languages such as C\# and Python are also in use.

\appendix
\newpage

\section{Acronyms used in this paper}
The following table has been generated from the on-line Gaia acronym list:
\newline\newline
\addtocounter{table}{-1}
\begin{longtable}{|l|p{0.8\textwidth}|}\hline 
\textbf{Acronym} & \textbf{Description}  \\\hline
ADASS&Astronomical Data Analysis Software and Systems \\\hline
AGIS&Astrometric Global Iterative Solution \\\hline
BSC&Barcelona Supercomputing Centre \\\hline
CERN&Centre Europ\'enne pour la Recherche Nucl\'eaire \\\hline
CESCA&CEntre de Supercomputac\'io de CAtalunya \\\hline
CNES&Centre National d'Etudes Spatiales (France) \\\hline
CPU&Central Processing Unit \\\hline
DPAC&Data Processing and Analysis Consortium \\\hline
DPACE&Data Processing and Analysis Consortium Executive \\\hline
ESA&European Space Agency \\\hline
ESAC&European Space Astronomy Centre (VilSpa) \\\hline
FLOP&FLoating-point OPeration \\\hline
GB&GigaByte \\\hline
HPC&High-Performance Computing \\\hline
ICRS&International Celestial Reference System \\\hline
IEEE&Institute of Electrical and Electronic Engineers \\\hline
JIT&Just In Time Compiler \\\hline
JNI&Java Native Interface \\\hline
JVM&Java Virtual Machine \\\hline
JWST&James Webb Space Telescope (formerly known as NGST) \\\hline
MB&MegaByte \\\hline
OO&Object Oriented \\\hline
PMD&Software tool to detect software problems \\\hline
RAM&Random Access Memory \\\hline
RMI&Remote Method Invocation \\\hline
SAN&Storage Area Network \\\hline
SOC&Science Operations Centre \\\hline
TB&TeraByte \\\hline
UB&University of Barcelona (Spain) \\\hline
UML&Unified Modeling Language \\\hline
VM&Virtual Machine \\\hline
\end{longtable}

\newpage
\section{Bibliography}
\bibliographystyle{spbasic}
\bibliography{womullan}

\end{document}